**Title**

Molecular Dynamics on Quantum Annealers

**Authors**

Igor Gayday,[1] Dmitri Babikov,[1*] Alexander Teplukhin,[2] Brian K. Kendrick,[3] Susan M. Mniszewski,[4] Yu Zhang,[3] Sergei Tretiak,[3] and Pavel A. Dub[5*]

**Affiliations**

[1] Department of Chemistry, Wehr Chemistry Building, Marquette University, Milwaukee, Wisconsin 53201-1881, USA

[2] Institute for Advanced Computational Science and Department of Chemistry, Stony Brook University, Stony Brook, New York 11794, USA

[3] Theoretical Division, Los Alamos National Laboratory, Los Alamos, New Mexico 87545, USA

[4] Computer, Computational and Statistical Sciences Division, Los Alamos National Laboratory, Los Alamos, New Mexico 87545, USA

[5] Chemistry Division, Los Alamos National Laboratory, Los Alamos, New Mexico 87545, USA

[*] Corresponding author, e-mail: dmitri.babikov@mu.edu

[*] Corresponding author, e-mail: pdub@lanl.gov

**Abstract**

In this work we demonstrate a practical prospect of using quantum annealers for simulation of molecular dynamics. A methodology developed for this goal, dubbed Quantum Differential Equations (QDE), is applied to propagate classical trajectories for the vibration of the hydrogen molecule in several regimes: nearly harmonic, highly anharmonic, and dissociative motion. The results obtained using the D-Wave 2000Q quantum annealer are all consistent and quickly converge to the analytical reference solution. Several alternative strategies for such calculations are explored and it was found that the most accurate results and the best efficiency are obtained by combining the quantum annealer with classical post-processing (greedy algorithm). Importantly, the QDE framework developed here is entirely general and can be applied to solve any system of first-order ordinary nonlinear differential equations using a quantum annealer.

**MAIN TEXT**

**Introduction**

Full-fledged quantum computers are expected to offer significant computational advantages compared to the traditional (classical) computers but, realistically, this is a distant future (*1*, *2*). Present-day quantum computers have relatively modest computational capabilities and are not yet free of technical issues such as noise and decoherence that cause errors (*3*). Still, these early machines can be used as a testbed for the development of quantum algorithms and for the small-scale proof-of-principle simulations, paving the way to practical applications on the next generation hardware. One of the long-thought applications of quantum computers is to simulate quantum systems, such as molecules and materials at the atomistic scale (*4–10*). Several algorithms were recently proposed for the calculations of molecular electronic structure (quantum chemistry) using either gate-based quantum computers with only a few tens of qubits, such as Sycamore (*11*, *12*), IBM Q-machines (*13*, *14*) or using quantum annealers with a larger number of qubits, such as the D-Wave systems (*15–17*). Another essential component of chemical modeling is the propagation of the equations of motion for constituent atoms in time and space (e.g., along the reaction path), called molecular dynamics simulations. Several steps have been made in this direction as well, including a method for solution of the quantum vibrational eigenvalue problem on a quantum annealer (*18*, *19*), and the trajectory simulation of molecular vibrations on the IBM quantum devices (*20*, *21*). These pioneering calculations were restricted to small molecules, such as diatomic and triatomic systems, and were often done in a hybrid quantum/classical fashion, when only a part of the overall algorithm is run on a quantum device. For example, in the trajectory simulations of Ref. (*20*), a quantum computer was used only to obtain the gradient of the molecular potential energy (the force determined by electrons), whereas the motion of the atoms (the actual trajectory) along this potential energy surface was propagated on a classical computer.

The goal of the present paper is to demonstrate that quantum architectures, in our case the D-Wave quantum annealer, can be used to simulate the molecular dynamics component pertaining to any computational chemistry problem, with possible future applications in chemical dynamics, material science or drug design. For this purpose, we developed a new method named Quantum Differential Equations (QDE) and carried out the first ever calculations of this sort on an actual quantum annealer (D-Wave 2000Q) to propagate trajectories for the motion of atoms in the hydrogen molecule, $H_2$ (i.e., vibrations). Although this is the simplest diatomic molecule, our method and code are general and suitable for simulation of polyatomic molecules with multiple vibrational modes, which will be accessible on the next generation of D-Wave annealers, such as Advantage (*22*). Moreover, our algorithm paves the way to the full-quantum molecular dynamics simulations within the time-dependent framework, including such methods as quantum trajectory, path-integral or thawed Gaussian (*23*).

**Results**

**Solution of Differential Equations on Quantum Annealers**

In general, the goal of molecular dynamics is to determine how a system of interacting atoms evolves in time. Here we will focus on the vibration of one molecular bond, when the trajectory is obtained by solving the following system of Hamilton's equations:

$$\begin{cases} \dfrac{dr}{dt} = \dfrac{p}{\mu} \\ \dfrac{dp}{dt} = F(r) \end{cases} \quad (1)$$

where $t$ is time, $r(t)$ is the bond length, $p(t)$ is its associated momentum, $\mu$ is a reduced mass, and $F(r)$ is the force field acting on atoms due to the local gradient of potential energy. In polyatomic molecules, equations similar to Eq. (1) can be written for every degree of freedom, coupled through the overall force field $F(r_1, r_2, r_3, \dots)$. Therefore, a general analogue of Eq. (1) can be written as the following system of $N$ differential equations:

$$\begin{cases} \dfrac{dy_1}{dx} = f_1(x, \bar{y}) \\ \quad \vdots \\ \dfrac{dy_N}{dx} = f_N(x, \bar{y}) \end{cases} \quad (2)$$

where $\bar{y} = (y_1, y_2, \dots, y_N)$ is a vector of $N$ unknown functions $y_n$, and $f_n$ is a set of $N$ known arbitrary functions. Without loss of generality, we can consider a system of 1st-order equations, since higher-order equations can always be rewritten as a system of 1st-order equations. To represent the functions $y_n$ and $f_n$ numerically, we introduce an equidistant grid of $M$ points with step size $\Delta x$ over the range of problem-specific values of the variable $x$. The functions $y_n$ and $f_n$ are then represented as arrays of their values at the grid points: $y_{n,i} = y_n(x_i)$ and $f_{n,i} = f_n(x_i, \bar{y}(x_i))$. Our goal is to formulate this problem as a binary minimization problem, which will enable its solution on a quantum annealer (25). While other methods to solve systems of differential equations on quantum annealers have also been considered in the literature (26, 27), we believe our approach is more general.

The only problem that a quantum annealer can handle is the minimization of a user-defined functional (subject to certain limitations, as discussed below). Therefore, to make system (2) solvable on a quantum annealer, we need to define a functional such that the correct solution vector $\bar{y}$ minimizes its value. One way to do this is to introduce a cost function $\epsilon(\bar{y})$ defined as the total squared difference between left- and right-hand sides of all equations in the system, at all points of the grid, namely:

$$\epsilon(\bar{y}) = \sum_{i=1}^{M} \sum_{n=1}^{N} \left( \left(\dfrac{dy_n}{dx}\right)_i - f_{n,i} \right)^2 \quad (3)$$

The value of the derivative $dy_n/dx$ can be approximated by the first order finite difference scheme:

$$\left(\dfrac{dy_n}{dx}\right)_i = \dfrac{y_{n,i+1} - y_{n,i}}{\Delta x} \quad (4)$$

Substituting this into Eq. (3) we obtain:

$$\epsilon(\bar{y}) = \sum_{i=1}^{M} \sum_{n=1}^{N} \dfrac{y_{n,i+1}^2 - 2 y_{n,i+1} y_{n,i} + y_{n,i}^2}{\Delta x^2} - 2 f_{n,i} \dfrac{y_{n,i+1} - y_{n,i}}{\Delta x} + f_{n,i}^2 \quad (5)$$

Note, that Eq. (5) treats all values of $y_n$ on an equal footing, which may not be desirable in practice, for example, if the relevant ranges of their values are substantially different. In this case, data rescaling or introduction of a penalty factor might be helpful to achieve better results.

The first practical consideration for Eq. (5) is that the variation of the solution at all points of the grid at once (one run with maximum value of $M$) may not be computationally feasible using present-day quantum annealers. Instead, one can consider only a subset of points in a given run of the annealer, and then use the final point obtained in this run as an initial condition for the next subset of points (smaller $M$, but multiple runs). In the limiting case when only one point of the grid is considered at a time (i.e. $M = 1$, maximum number of consecutive runs), the values of $f_{n,i}$ are all known from the previous step, which decouples all equations (5) and makes it possible to solve them one by one. This procedure permits us to solve, piece by piece, even large systems, even on annealers with a small number of qubits. As more qubits become available in the future, they can be effectively utilized to consider more grid points at the same time, and thus reduce the total number of runs.

Another practical problem is that $f_n(x, \bar{y})$ is an arbitrary function, which makes Eq. (5) non-quadratic in general, whereas the current generation of quantum annealers does not have native support for non-quadratic functions. To keep Eq. (5) quadratic in $\bar{y}$, the $f_n(x, \bar{y})$ has to be no more than linear in $\bar{y}$. One way to solve this issue is to split $f_n(x, \bar{y})$ into linear segments and find the solution for each segment separately, using the last solution point in each segment as an initial condition for the next one. This separation into multiple runs is similar to what has been discussed in the previous paragraph and both of these procedures can be combined. Namely, when solving for only a few grid points per run (so that the solution does not advance too far), the local behavior of $f_n(x, \bar{y})$ around the initial point of this run can be approximated with a linear segment reasonably well. In the case when only one point is considered at a time, the values of $f_{n,i}$ are constant, so the issue of separating into linear segments does not arise at all.

Now let us obtain an expression for the functional $\epsilon(\bar{y})$ at each segment. The right-hand side function $f_n(x, \bar{y})$ is linear with respect to $\bar{y}$ within a given segment, so it can be written as:

$$f_n(x, \bar{y}) = f_{n,0}(x) + \sum_{k=1}^{N} f_{n,k}(x) y_k \tag{6}$$

where the index $k$ labels individual terms in the linear expansion of $f_n(x, \bar{y})$. Plugging this into Eq. (5) and expanding squared terms, one obtains the following expression (quadratic in $\bar{y}$):

$$\epsilon(\bar{y}) = \sum_{i=1}^{M} \sum_{n=1}^{N} \frac{y_{n,i+1}^2 - 2y_{n,i+1} y_{n,i} + y_{n,i}^2}{\Delta x^2} - 2\left(f_{n,0,i} + \sum_{k=1}^{N} f_{n,k,i} y_{k,i}\right)\left(\frac{y_{n,i+1} - y_{n,i}}{\Delta x}\right)$$

$$+ \left(f_{n,0,i}^2 + 2f_{n,0,i}\left(\sum_{k=1}^{N} f_{n,k,i} y_{k,i}\right) + \sum_{k=1}^{N} \sum_{k'=1}^{N} f_{n,k,i} f_{n,k',i} y_{k,i} y_{k',i}\right) \tag{7}$$

From this expanded form, it is easy to see the individual coefficients for all unknown variables and to compose the so-called Quadratic Programming (QP) (28) matrices $H$ and $d$ for this function, such that Eq. (7) can be written in matrix form as:

$$\epsilon(\bar{y}) = \bar{y}^T H \bar{y} + \bar{y}^T d \tag{8}$$

## Adaptation to Binary Variables

The last step that we need to do before we can submit this task to a quantum annealer is to convert the QP-matrices into the corresponding QUBO matrix $Q$ (for Quadratic Unconstrained Binary Optimization) (25), which has the same meaning as QP-matrices, but, instead of the continuous variables of Eqs. (7) and (8), it uses binary variables $q_{n,i,j} = \{0,1\}$ that represent the states of the qubits after the read operation. This conversion can be done as follows. First, a given continuous variable $y_{n,i}$ can be approximated by a set of binary variables $q_{n,i,j}$ via a regular signed fixed-point number representation (18):

$$y_{n,i} = -2^{K_I-1} + \sum_{j=-K_I+1}^{K_D} 2^{-j} q_{n,i,j} \tag{9}$$

where $K_I$ and $K_D$ are discretization parameters, equal to the number of qubits used to represent integer and decimal parts of $y_{n,i}$, respectively. Equation (9) can approximate numbers in the range from $-2^{K_I-1}$ to $2^{K_I-1} - 2^{-K_D-1}$ with an error of up to $2^{-K_D-1}$. The floating-point representation, while being superior to the fixed-point representation on classical computers, is not used here because this representation is not quadratic in $q_{n,i,j}$, which makes it unsuitable for use on quantum annealers, as discussed above.

Next, to express the functional $\epsilon(\bar{y})$ in terms of the binary variables $q_{n,i,j}$, one could plug Eq. (9) into Eq. (7), and compose the QUBO-matrix $Q$ in the same way the QP-matrices $H$ and $d$ were composed. But this is tedious, and the resulting expression is rather long. Therefore, we will not derive it explicitly here. Instead, the QUBO-matrix can also be obtained directly from the QP-matrices by simply replacing each element of the matrix $H$ with a $K \times K$ block ($K = K_I + K_D$), where the value of element $(i, i', j, j')$ is given by:

$$Q_{ii'jj'} = \begin{cases} 2^{-2j} H_{ii} + 2^{-j} d_i - \sum_k 2^{K_I-1-j}(H_{ik} + H_{ki}), & \text{if } i = i' \text{ and } j = j' \\ 2^{-(j+j')} H_{ii'}, & \text{otherwise} \end{cases} \tag{10}$$

Here $i$ and $i'$ are indices of rows and columns of the original QP-matrices $H$ and $d$, while $j$ and $j'$ are local indices of rows and columns within each $K \times K$ block. The range of $j$ in each block is the same as in Eq. (9): from $-K_I + 1$ to $K_D$. Note that since $q_{n,i,j} = q_{n,i,j}^2$ there is no need for a separate vector $d$, as the relevant contributions are simply added to the diagonal of $Q$. Thus, the minimized functional $\epsilon$ can be written in QUBO formalism as:

$$\epsilon(\bar{q}) = \bar{q}^T Q \bar{q} \tag{11}$$

The QUBO-matrix defined by Eq. (10) can be used as a direct input to a quantum annealer to obtain the values of $\bar{q}$ that minimize Eq. (11), which, in turn, can be used to reconstruct the continuous solution to the problem of Eq. (2).

Practical implementations of this method were carried out using the D-Wave 2000Q quantum annealer at Los Alamos National Laboratory. The values of molecular parameters were chosen to represent H₂ with an analytical Morse potential as described in the *Supplemental Information* (in principle, an arbitrary potential energy function could be used). Discretization parameters $K_I$ and $K_D$ were set to 6 and 15, respectively. This choice can represent the numbers from –32 to ≈ +32 with rounding errors up to $2^{-16} \approx 10^{-5}$ (in atomic units for both coordinate and momentum), using $K_I + K_D = 21$ qubits per number. The D-Wave 2000Q has 2048 physical qubits, but each qubit can only interact with 6 neighbors. In cases when full connectivity is required, a number of physical qubits is treated as a single logical qubit. The maximum number of such fully connected logical qubits on D-Wave 2000Q is 64 (22), which is sufficient to either propagate through one time step using both equations of the system of Eq. (1) at once, or to propagate the equations for position and momentum individually one after another. We tested both strategies. All trajectories in this work were propagated through the time interval close to 10 femtoseconds (400 a.u. of time, or about 1.25 vibrational periods of the ground state H$_2$). The only exception are the trajectories in **Fig. 2**, where the propagation time was doubled.

## Example Trajectories

**Figure 1** shows several solutions for a low energy trajectory which starts at $r_0 = 1.3$ Bohr with zero initial momentum ($p_0 = 0$), all of which are computed with 1000 equal time steps, i.e. $\Delta t \approx 10$ attoseconds. This bond length is close to the equilibrium position, so the vibrational motion is expected to be fairly harmonic. The black dashed line in **Fig. 1** shows the exact analytical solution available for this problem (see the *Supplemental Information* for details). The red line shows a direct application of our method, using the D-Wave quantum annealer to propagate one equation at a time through one time step ($N = 1, M = 1$). As one can see, the accuracy of this trajectory is rather poor. One way to improve the quality of D-Wave's results is to simply restart the calculation of each trajectory point multiple times and select the best solution among these runs (i.e., the one with the smallest $\epsilon$) as the answer. The effect of this strategy is shown with the green line, where each task was allowed to be solved up to 10 times. Comparing these two trajectories, one can see that restarting improves the quality of the D-Wave's results quite dramatically.

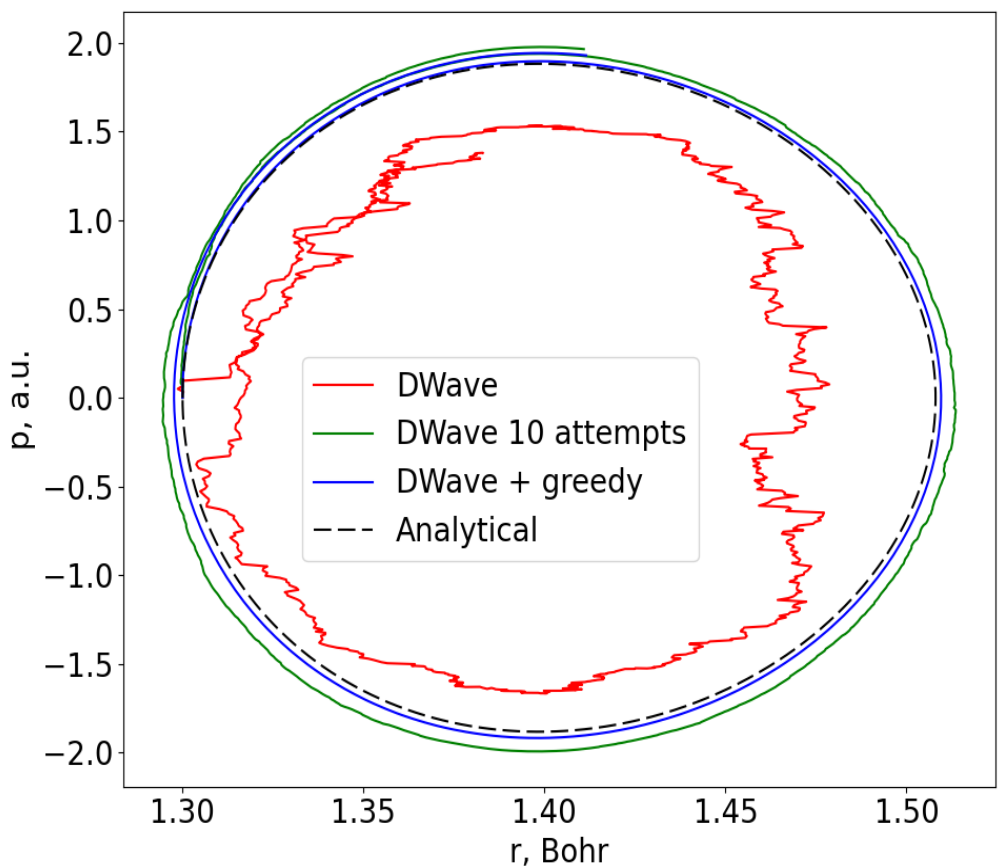

**Fig. 1: Trajectory for the vibration of the H$_2$ molecule in the low-energy regime of nearly harmonic motion.** The trajectory is plotted in the phase space (momentum vs. coordinate). The points are connected in the order of time and start from $r_0 = 1.3$ Bohr with $p_0 = 0$. The exact analytical solution (dashed black) and three solutions obtained on a quantum annealer (color lines) are presented.

Another way to improve the accuracy is to allow some classical postprocessing. The blue line in **Fig. 1** shows an example of a trajectory where each solution found by D-Wave was used as a starting point for a classical greedy algorithm (available as a part of D-Wave's Ocean tools),(24) which then tried to improve the solution found by the quantum annealer. Both equations were propagated at once in this case ($N = 2, M = 1$) and only one attempt per job was allowed, but this was sufficient to obtain an even better trajectory. We conclude that this hybrid quantum-classical approach is the most efficient and accurate across the three scenarios considered here.

The next set of results is presented in **Fig. 2**, where we increased the time step ten-fold (i.e., $\Delta t \approx 100$ attoseconds). In this picture each point of the time-grid is shown by a symbol, to show the discrete nature of the numerical solution. In addition to the analytical and the D-Wave's results, we also present a trajectory obtained with the QP version of our algorithm, by minimizing the target functional in the form of Eq. (8) on a classical computer. This method is free of the technical issues that are still present in the D-Wave annealer. In this case it serves as the best solution that one can achieve within the framework of our QDE algorithm at a chosen level of discretization, if all minimizations are carried out correctly. **Figure 2A** represents the same trajectory as **Fig. 1** and shows that increasing the time step by a factor of 10 leads to a noticeable deviation of all numeric solutions (including QP) from the exact analytic solution, but the results obtained by the D-Wave quantum annealer remain very close to the QP result. We also tried to propagate trajectories with the high level of vibrational excitation, when the motion is expected to be highly anharmonic. **Figure 2B** illustrates one example of such a trajectory, where the energy of $H_2$ was close to 50% of its dissociation energy ($r_0 = 0.90$ Bohr, which corresponds to a compressed bond, $p_0 = 0$). **Figure 2C** gives another example, where the energy of $H_2$ is chosen above the dissociation threshold and this leads to a relatively fast bond-breaking, or dissociation of the molecule into atoms ($r_0 = 0.70$ Bohr in the repulsive range, $p_0 = 0$). The level of discretization and the methods/colors are the same as in **Fig. 2A**. We see that the results of a combined quantum-classical approach (D-Wave + greedy) overlap with the QP results in all three frames of **Fig. 2**.

When comparing **Figs. 1** and **2**, one can see how multiple components of the overall error influence the quality of solutions. In **Fig. 2**, the overall error is dominated by the propagation error of QDE discretization (i.e., large time step), so the internal D-Wave's error (i.e., error of minimization of the target functional) is not noticeable and both QP and D-Wave's solutions look similar. In contrast, in **Fig. 1** the time step is small, so the propagation error becomes negligible, and the internal D-Wave's error becomes apparent, which is responsible for the "ragged" look of the D-Wave's solution in this case.

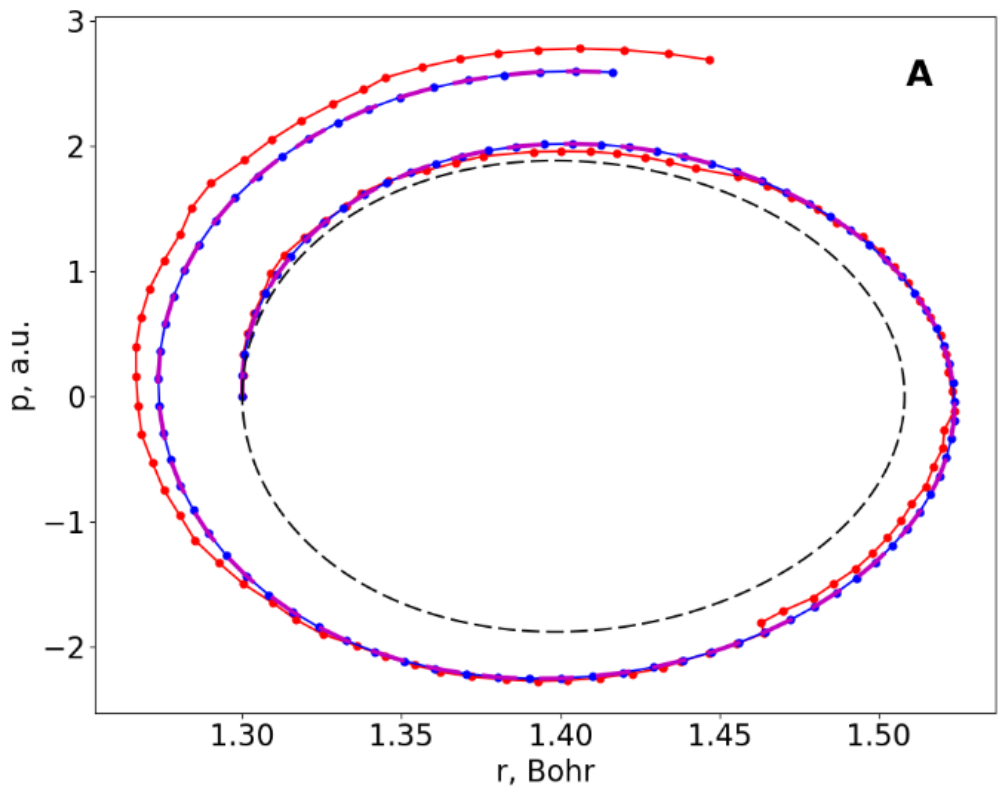
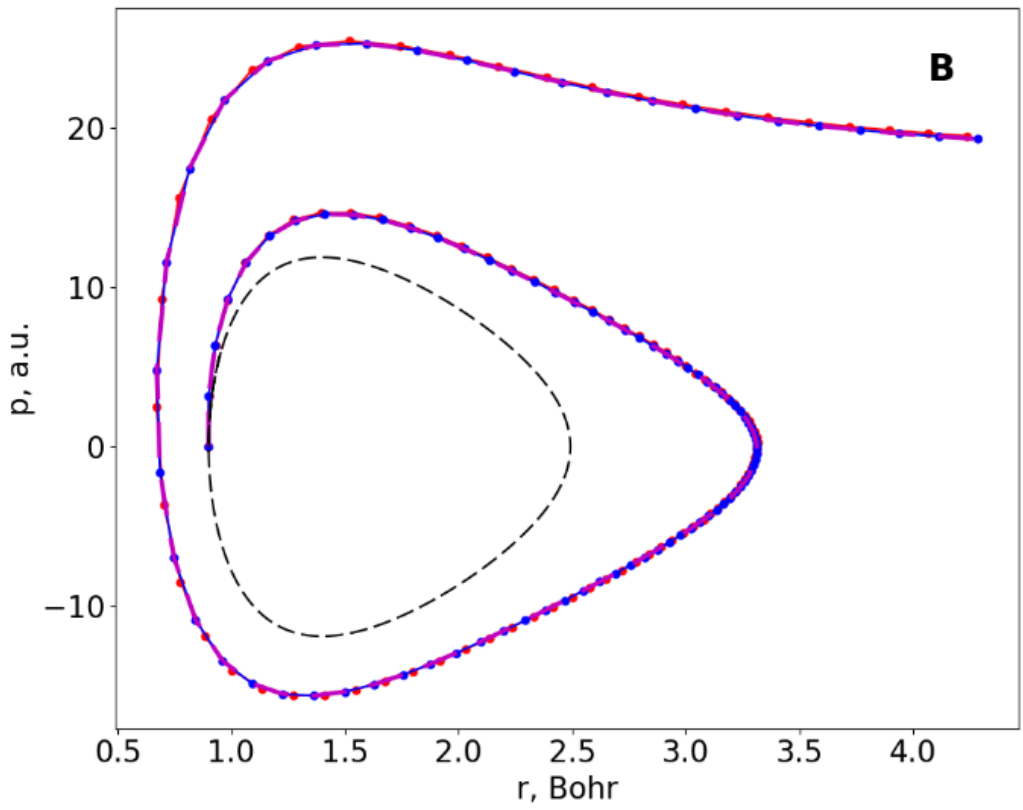

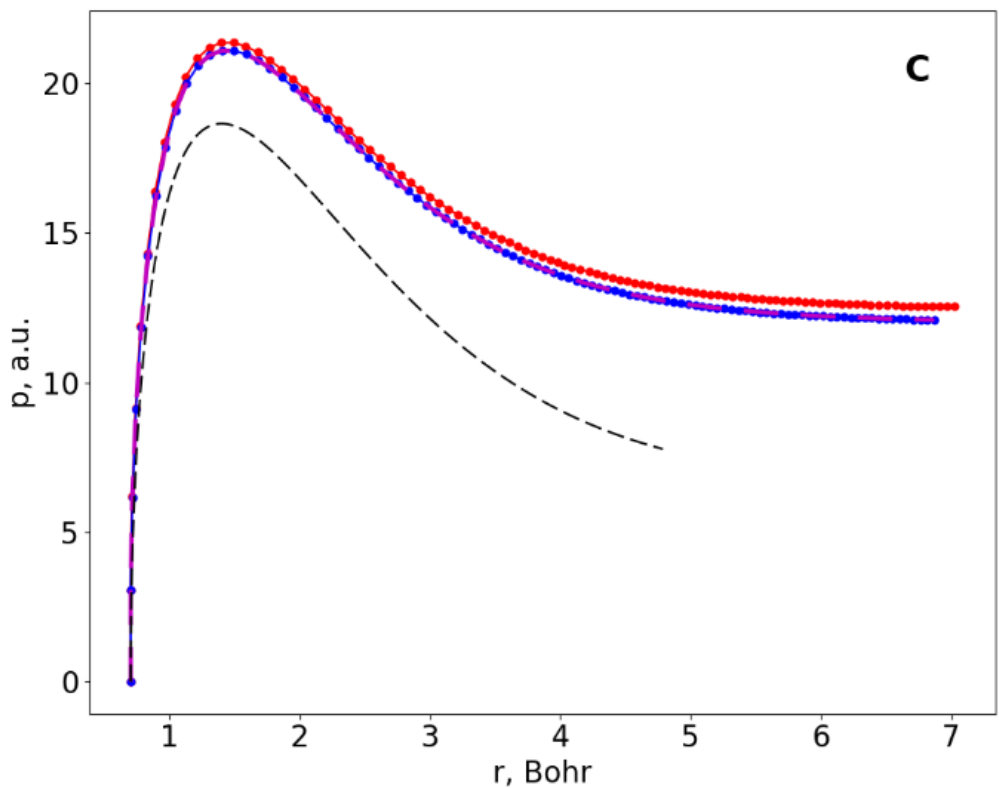

**Fig. 2: Different types of trajectories for the vibration of the H$_2$ molecule.** All trajectories are plotted in the phase space (momentum vs coordinate): **(A)** in the low-energy regime of nearly harmonic motion, **(B)** in the high-energy regime of strongly anharmonic motion, and **(C)** at an energy above the dissociation threshold that leads to bond breaking. The exact analytical solution (dashed black), a classical QP solution (dashed magenta) and solutions obtained using the D-Wave quantum annealer with (blue) and without (red) greedy postprocessing are presented for each case.

## Convergence Analysis for N = 1

To provide more insight into the performance of the QDE method itself, and of its execution on the D-Wave annealer, we have analyzed these trajectories on a more quantitative level. In **Fig. 3** we demonstrate convergence properties for several approaches we tried, by plotting their values of Residual Mean Squared Error (RMSE, see the *Supplemental Information*) with respect to the analytical trajectory, as a function of the total number of grid points or time steps (larger number of grid points corresponds to smaller time step $\Delta t$). All data in **Fig. 3** were obtained using the simplest version of the QDE algorithm ($N = 1, M = 1$, when one equation is propagated at a time through one time step). In these and further tests, the value of $r_0$ was set to 1.3 Bohr and $p_0 = 0$, the same as in **Fig. 1**.

First, we solved Eq. (1) classically, as a continuous QP-problem (magenta line in **Fig. 3**). Once again, the QP-solution provides a helpful reference because it is exact and shows the minimum error that can be achieved for a given number of grid points, if the function of Eq. (8) is minimized correctly. One can see that our method converges quickly with respect to the total number of grid points, which is the only convergence parameter that we have here. Next, we found solutions in binary variables (similar to qubits) using the QBSolv software tool also available as a part of D-Wave's Ocean tools (*24*), which uses a classical heuristic probabilistic algorithm to find the minimum of Eq. (11) in binary variables. Finding the global minimum in discrete variables is harder than in continuous variables, therefore, in contrast to the QP solver, QBSolv is not guaranteed to find the correct solution. However, as one can see from **Fig. 3**, the solutions obtained with QBSolv (dashed line) nearly coincide with the ideal QP-solutions, so this algorithm is working well for small QUBO-matrices ($K_I + K_D = 21$). This test demonstrates that the correct solution to Eq. (1) can be obtained in binary variables if minimization of Eq. (11) is done sufficiently accurately.

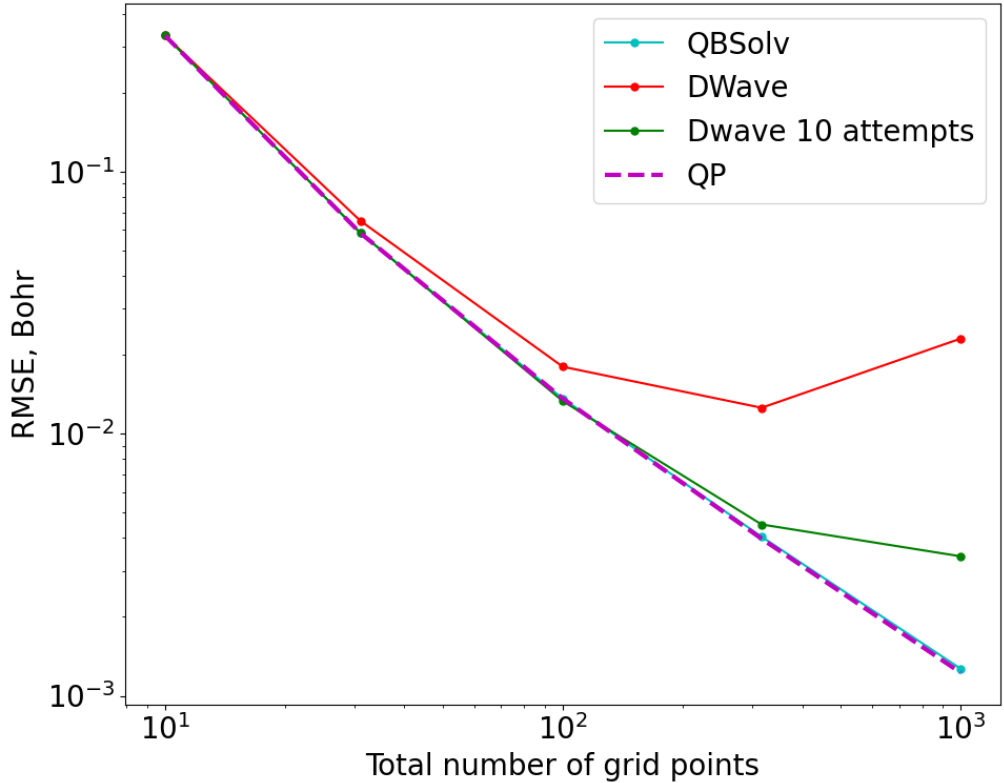

**Fig. 3: RMSE of solutions obtained with different methods as a function of the total number of grid points (or time steps).** All methods propagate one equation at a time.

As a next step, we used the D-Wave to minimize Eq. (11), instead of the classical solver QBSolv. As one can see from the red line in **Fig. 3** (which also corresponds to the red line in **Fig. 1** for 1000 grid points), the D-Wave's result is close to the results of QP and QBSolv at small values of $M$, but deviates further at larger values of $M$, unable to achieve low values of RMSE. Loosely speaking, we define an "accurate" trajectory as one with RMSE < $10^{-2}$ Bohr, which means that in this test the D-Wave failed to provide any acceptable solution. From the QBSolv's result we know that better solutions exist, but D-Wave's annealer was not able to find them. We can conclude that the D-Wave has internal errors or noise that results in an RMSE on the order of $10^{-2}$ Bohr (for our problem), and is responsible for the deviation of the red line observed in **Fig. 3**. We tried to improve this result by varying annealing parameters, such as annealing time and chain strength (*25*), but this did not lead to any significant changes in the errors of trajectories. The only parameter that reliably improved the quality of solutions in this test was the number of reads, which was set to the maximum allowed value of 10000 for all D-Wave's results in this work.

Finally, the green line in **Fig. 3** corresponds to the green line in **Fig. 1**, i.e., to the case when each point was restarted multiple times (up to 10 attempts). This may be similar to increasing the number of reads beyond the maximum value, depending on the D-Wave's internal implementation of it. We see that this simple fix reduces the RMSE of the solutions by almost an order of magnitude, and produces a sufficiently accurate trajectory.

## Convergence Analysis for N = 2

In the next computational experiment, reported in **Fig. 4**, we tried to propagate the equations for position and momentum together in one run ($N = 2$). Once again, the magenta line shows perfect results of the QP solver, which can be used as a reference line. The results of QBSolv with 1 attempt per task (cyan line) in this case are significantly worse, but this is expected since the algorithm is classical, and the size of the search space increases exponentially with the size of QUBO matrix. However, similar to what we saw for D-Wave in **Fig. 3**, the results of QBSolv can be significantly improved simply by restarting each task several times (yellow line) and improved even further if one rescales the ranges of two variables in the two equations (brown line) to equalize contributions of position and momentum to the minimization functional (see the *Supplemental Information*). In this final form, the accuracy of QBSolv approaches that of QP.

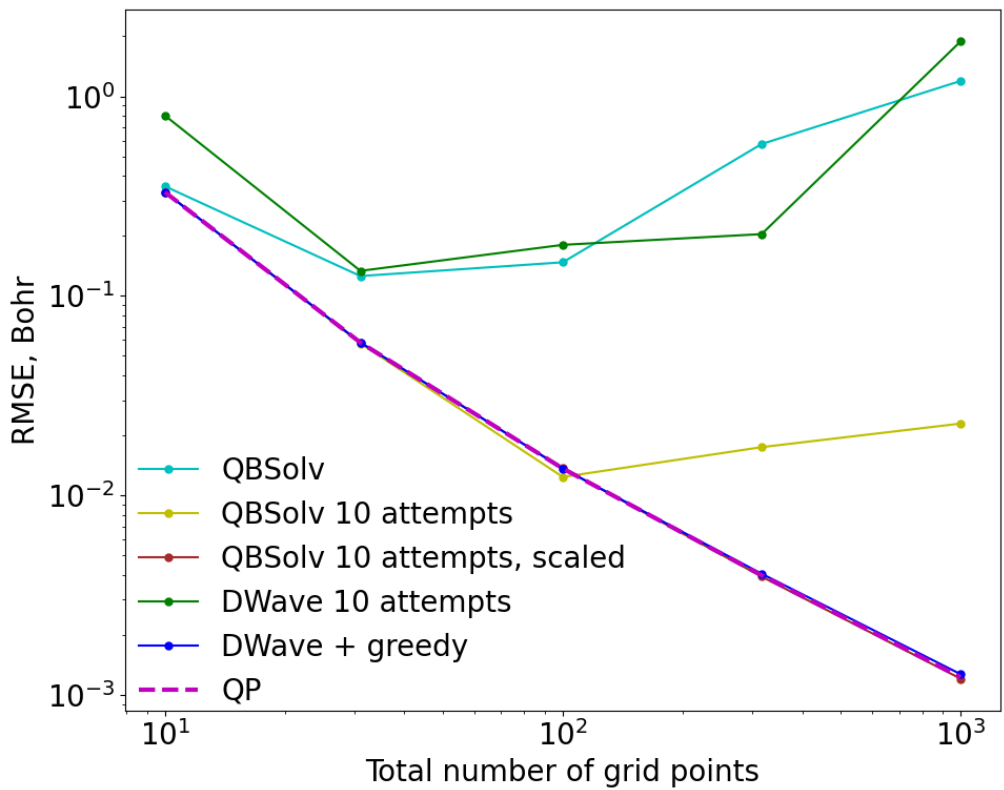

**Fig. 4: RMSE of solutions obtained with different methods as a function of the total number of points in the grid.** All methods propagate both equations at once.

Returning back to the performance of the D-Wave annealer, we see that even with 10 attempts per task (green line in **Fig. 4**) the results are much worse compared to the analogous results for the case when only one equation was propagated at a time (green line in **Fig. 3**). This is in sharp contrast with classical QBSolv and could indicate that the overall quality of the quantum annealing hardware is reduced when the number of qubits is increased for a given problem. We suspect that such dependence could exist due to the fact that only a few other qubits are directly connected to any given qubit on the D-Wave and involvement of a larger number of qubits translates into longer qubit connection chains, which are imperfect. Surprisingly, rescaling the variables in the two equations did not help in this case either.

However, we found that a hybrid strategy (when each solution found by D-Wave was used as a starting point for the classical greedy algorithm) works really well for this harder problem (blue line in **Fig. 4**, which also corresponds to blue line in **Fig. 1**). This result matches the quality of QP solutions even for a large number of grid points (i.e., small time steps and high accuracy). One might think that the greedy algorithm is doing all the work here and that the involvement of D-Wave is unnecessary. In order to rule out this possibility, we tried to lower the quality of D-Wave's solutions by significantly reducing the number of reads, which basically gave a near-random starting point for the greedy algorithm. We found that in this case greedy algorithm does not perform well, which indicates that the D-Wave's initial guess is important here. Notably, while the purely classical algorithm (QBSolv) was also able to achieve the same results, the hybrid quantum-classical strategy (D-Wave + greedy) did it much more easily with only one attempt per task and even without rescaling, which is quite optimistic. One explanation of the success of this hybrid approach is that D-Wave is good at identifying the global minimum, but it cannot descend all the way there with a high precision due to the presence of noise and related errors. On the other hand, greedy algorithm is good at descending to the very bottom of the minimization functional but needs to be placed in the vicinity of the global minimum first. Thus, the two methods complement each other and together converge to a very good result.

**Discussion**

Practicality of presented numerical algorithms is obviously defined by the scaling of hardware requirements with the system size. Overall, the resources required by QDE grow as $O(MN)$, where $M$ is the total number of points in the grid, which depends on the desired accuracy, and $N$ is the total number of equations in the system, which increases linearly with the dimensionality of the problem. For example, the new D-Wave Advantage quantum annealer with 177 fully connected logical qubits (*22*) could be used to propagate up to 8 equations in one run, which is sufficient to describe the vibrational motion of any triatomic and some tetratomic molecules. For larger (polyatomic) molecules the trajectories could be propagated by splitting the overall workload into several runs, as described above.

In cases when the right-hand side functions $f_n(x, \bar{y})$ of Eq. (2) are non-linear, our method will not be able to obtain all points of a solution at once, even on annealers with sufficiently large number of qubits, since only quadratic functions can be minimized on such devices. As explained above, this can be easily circumvented by running the same algorithm multiple times. Ultimately, the minimum required number of consecutive runs will be dictated by the number of segments approximating $f_n(x, \bar{y})$, which depends on desired accuracy of a solution and the degree of non-linearity of the $f_n(x, \bar{y})$. However, we can still make use of all available resources of a given (large, if available) annealer by increasing the number of grid points in each segment. Note that this restriction only concerns the dependence of $f_n(x, \bar{y})$ on $\bar{y}$; the dependence on $x$ can still be arbitrary and cause no problems. Although the methods to simulate higher-order polynomial behavior on quantum annealers exist and can be used to approximate $f_n(x, \bar{y})$ with segments of higher order to reduce the total number of necessary runs, this does not necessarily work better for arbitrary functions and was not discussed here.

To summarize, in this work we demonstrated that quantum annealers can be used for molecular dynamics simulation, which constitutes an essential component of the computational chemistry and materials modeling "toolbox". The new methodology developed for this goal (the QDE method) was applied to propagate classical trajectories simulating vibrational motions of the hydrogen molecule (H$_2$) in three different energy regimes (nearly harmonic, highly anharmonic, and dissociative limits). The results obtained using the D-Wave 2000Q quantum annealer are mutually consistent and quickly converge to the analytical solution. Several alternative strategies for such calculations were explored and it was found that the most accurate results and the best efficiency are obtained by combining the quantum annealer with classical post-processing (greedy algorithm). Importantly, the QDE framework developed here is entirely general and can be applied to solve any system of first-order ordinary nonlinear differential equations using a quantum annealer. The new generation of quantum annealers, such as the D-Wave Advantage with more qubits and better connectivity, could be used to either compute multiple time-grid points at once ($M > 1$) or to explore more complicated molecules with many degrees of freedom ($N > 2$). Higher order finite difference schemes can also be useful in some applications. Development of trajectory based quantum algorithms, suitable for execution on a quantum annealer, is another promising potential future direction.

**Acknowledgments**

**Funding:**

MolSSI Investment Fellowship, funded by National Science Foundation grant ACI-1547580 (IG)

National Science Foundation grant CHE-2102465 (DB)

Research at Los Alamos National Laboratory (LANL) is supported by Laboratory Directed Research and Development (LDRD) program, project number 20200056DR and performed in part at the Center for Integrated Nanotechnologies (CINT), a U.S. Department of Energy, Office of Science user facility at LANL.

**Author contributions:**
  Conceptualization: BKK, PAD
  Methodology: IG, DB, AT, SMM
  Software: IG
  Investigation: IG, BKK, SMM
  Resources: BKK, PAD
  Writing—original draft: IG, DB
  Writing—review & editing: AT, BKK, SMM, YZ, ST, PAD
  Visualization: IG
  Supervision: DB, ST, PAD
  Funding acquisition: IG, DB, ST, PAD

**Competing interests:** Authors declare that they have no competing interests.

**Data and materials availability:** All data are available in the main text or the supplementary materials.

# Supplementary Materials for

## Molecular Dynamics Simulations on Quantum Annealers

Igor Gayday, Dmitri Babikov,[*] Alexander Teplukhin, Brian K. Kendrick, Susan M. Mniszewski, Yu Zhang, Sergei Tretiak, and Pavel A. Dub[*]

*Corresponding author. Email: dmitri.babikov@mu.edu
*Corresponding author. Email: pdub@lanl.gov

**This PDF file includes:**

  Supplementary Text
  Figs. S1 to S2

**Other Supplementary Materials for this manuscript include the following:**

  Python code that implements the method, application and figures discussed in this work
  All related data files

**Supplementary Text**

Analytical expressions for potential energy and force fields

The analytical Morse potential energy surface used in this work is given by
$$V(r) = D_e \left( e^{-2a(r-r_e)} - 2e^{-a(r-r_e)} \right) \tag{S1}$$

The parameters of this potential were adjusted to represent a hydrogen molecule, namely $D_e \approx$ 36450 cm$^{-1}$, $a \approx 1.04$ Bohr$^{-1}$, $r_e \approx 1.40$ Bohr, where $D_e$ is the dissociation energy, $a$ is the Morse constant (steepness of potential), and $r_e$ is the equilibrium position (lowest potential energy point). The resulting potential is shown in **Fig. S1**.

The force field derived from this potential is given by:
$$F(r) = -\frac{dV(r)}{dr} = 2aD_e \left( e^{-2a(r-r_e)} - e^{-a(r-r_e)} \right) \tag{S2}$$

Analytical solution to equations of motion in a Morse potential

The system of Eq. (1) in the main text can be re-written as a single second order differential equation:
$$\frac{d^2 r}{dt^2} = \frac{F(r)}{\mu} \tag{S3}$$

Analytical solution to Eq. (S3) for initial position $r_0$ and zero initial speed can be written as:
$$r(t) = \frac{1}{a} \ln \left( c_1^2 \tau \frac{c_3 D_e + \left( D_e - \frac{c_4}{\tau} \right)^2}{2 c_1 c_3 c_4} \right) \tag{S4}$$

where
$$c_1 = e^{a r_e} \tag{S5}$$
$$c_2 = e^{a r_0} \tag{S6}$$
$$c_3 = -D_e \frac{c_1}{c_2} \left( 2 - \frac{c_1}{c_2} \right) \tag{S7}$$
$$c_4 = D_e + \frac{c_2 c_3}{c_1} \tag{S8}$$
$$\tau = e^{\sqrt{\frac{2 c_3}{\mu}} a t} \tag{S9}$$

Note that this equation only provides us with the value of coordinate, so momentum needs to be calculated separately: either numerically (e.g. finite difference) or analytically from conservation of energy.

RMSE

The analytical solution of Eq. (S4) was used to calculate RMSE for a given solution $\hat{r}(t)$ as:

$$\text{RMSE} = \sqrt{\frac{1}{M}\sum_{i=1}^{M}(\hat{r}_i - r_i)^2} \quad (S10)$$

Performance of greedy with poor D-Wave answers

**Fig. S2** shows what happens when the quality of D-Wave's initial guess in the hybrid D-Wave + greedy approach is deliberately lowered by reducing the total number of reads to 10. Comparing these very wrong trajectories with the perfect results of D-Wave + greedy in the main text, where the number of reads was 10000, one can see that the quality of D-Wave's initial guess is important and D-Wave's involvement is essential.

Details of data rescaling

Rescaled results in Figure 4 of the main text were obtained by dividing momentum by 20 and shifting coordinate by -1.4 Bohr.

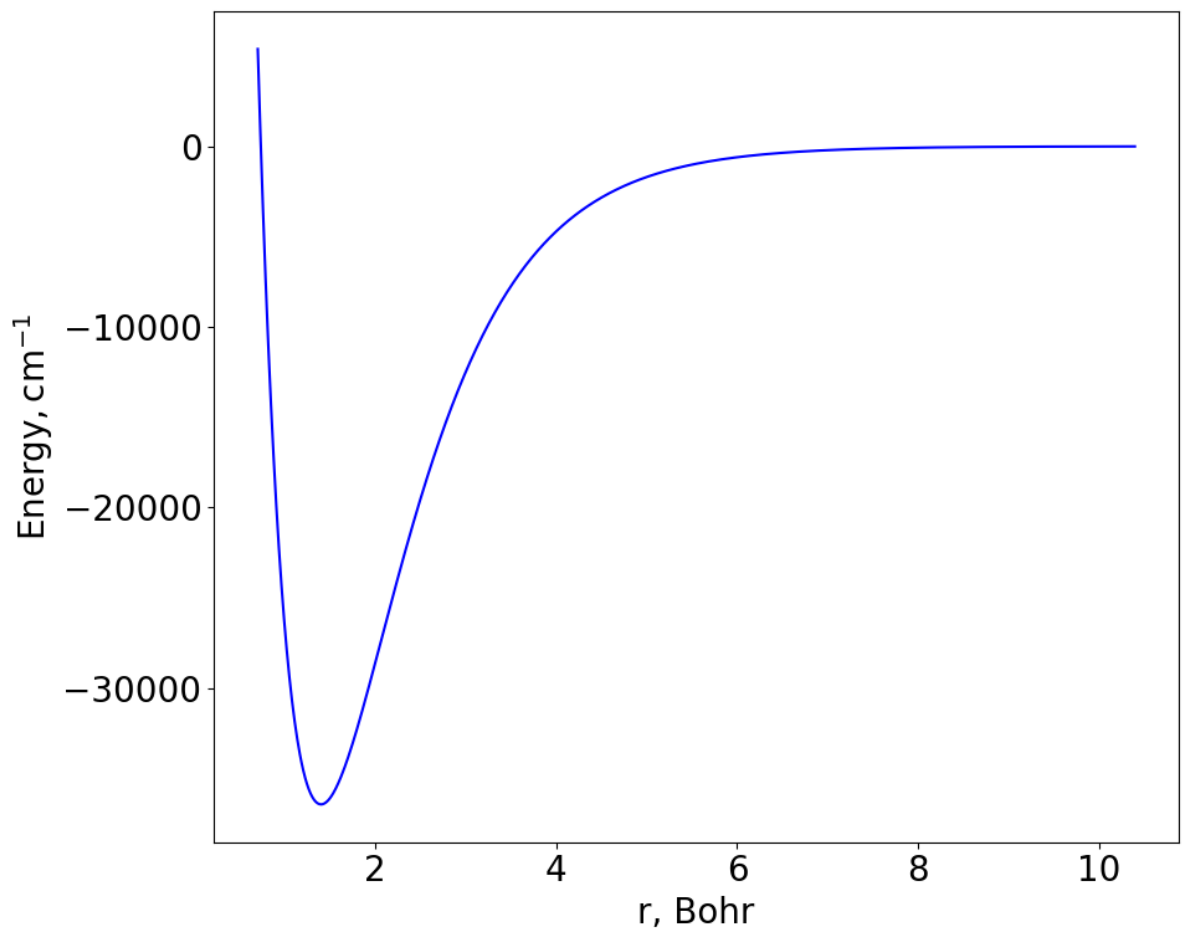

**Fig. S1 Potential energy surface used in this work.**

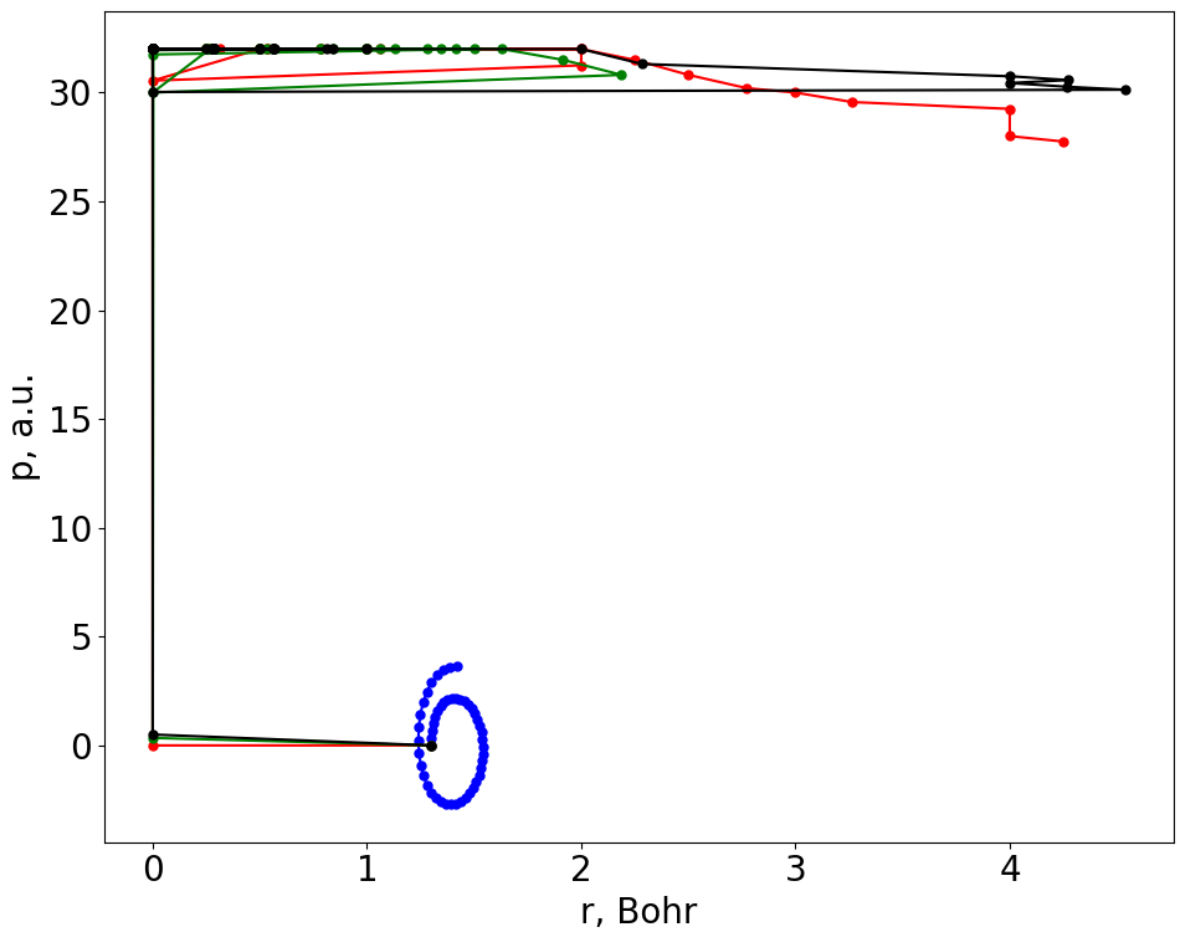

**Fig. S2: Several examples of trajectories obtained with hybrid D-Wave + greedy method, when the quality of D-Wave's answers was deliberately lowered.** The blue line shows ideal QP trajectory for comparison. The remaining three lines are three attempts to solve with D-Wave + greedy.

**Data S1. (separate file)**

The Python code that implements the method, application and figures discussed in this work, as well as all related data, is attached in an archive. Additionally, the code can be found at https://github.com/IgorGayday/qde/